\documentclass{article}
\usepackage{spconf,amsmath,amssymb,graphicx}
\usepackage{mathtools}
\usepackage{caption,subcaption,indentfirst,placeins}
\usepackage{color}
\usepackage{cite}

\title{Enhancing temporal segmentation by 
 nonlocal self-similarity}
\name{Mariella Dimiccoli$^{(1)}$ and Herwig Wendt$^{(2)}$}
\address{%
$^{(1)}$ Institut de Rob\`otica i Inform\`atica Industrial, CSIC-UPC, Barcelona, Spain.  {\tt\small mdimiccoli@iri.upc.edu}\\
$^{(2)}$ IRIT, CNRS, University of Toulouse, Toulouse, France. 
}
\graphicspath{{images/}}

\begin{document}
%
\maketitle
\begin{abstract}
Temporal segmentation of untrimmed videos and photostreams is currently an active area of research in computer vision and image processing. This paper proposes a new approach to improve the temporal segmentation of photostreams. The method consists in enhancing image representations by encoding long-range temporal dependencies.  Our key contribution is to take advantage of the temporal stationarity assumption of photostreams for modeling each frame by its nonlocal self-similarity function. 
The proposed approach is put to test on the EDUB-Seg dataset, a standard benchmark for egocentric photostream temporal segmentation. 
Starting from seven different (CNN based) image features, the method yields consistent improvements in event segmentation quality, leading to an average increase of F-measure of $3.71\%$ with respect to the state of the art. 
\end{abstract}
\section{Introduction}
\label{sec:intro}

With the proliferation of wearable and smartphone cameras in recent years,  the amount of untrimmed videos on internet is increasing exponentially. Consequently, we have witnessed a growing interest in developing algorithms to segment long unstructured videos into meaningful and manageable semantic units, commonly called \textit{events} \cite{zelnik2001event,liwicki2015online,poleg2014temporal,dimiccoli2017sr,del2018predicting}. Event segmentation is crucial not only to video understanding but also to video browsing, indexing and summarization. Additionally, the temporal segmentation of human motion into actions is central to the understanding and building of computational models of human motion and activity recognition \cite{spriggs2009temporal,kruger2017efficient}. Beside image data, time series segmentation is a core problem in data mining and machine learning with applications in several domains ranging from land cover changes tracking from  remotely-sensed data \cite{kennedy2010detecting,jamali2015detecting} to health monitoring with wearable sensor data streams  \cite{adams2016hierarchical,bari2018rconverse}, to name but a few.

Roughly speaking, a temporal segmentation algorithm consists of a feature extraction step followed by the segmentation process itself that acts on the extracted features \cite{koprinska2001temporal,abdulhussain2018methods} to detect transitions between shots. Typically, a crucial component of the temporal segmentation algorithm is a measure of dissimilarity/similarity among the extracted features.  

This paper focuses on the temporal segmentation of egocentric photostreams captured by a wearable photo camera. Given the very low frame rate (2fpm), these image sequences often present abrupt appearance changes even in temporally adjacent frames that harden the task of temporal segmentation (see Fig.\ref{fig:example}). State of the art approaches have focused on the segmentation algorithm \cite{dimiccoli2017sr}, or at improving the representation \cite{del2018predicting,paci2016context,dias2019learning} by learning approaches.
The aim of this paper is to explore the use of \emph{nonlocal self-similarity} for temporal segmentation \footnote{
Code available at: https://github.com/mdimiccoli/Nonlocal-self-similarity-1D}. We show that it allows to capture long-range temporal dependencies over the entire sequence that, whatever are the initial features used to represent single images, leads to improved temporal segmentation performance.


The remainder of this paper is organized as follows. Section \ref{sec:related-work} provides an overview over related work, while Sections \ref{sec:methodology} and \ref{sec:results} are devoted to detail the proposed methodology and to report and discuss experimental results, respectively. Section \ref{sec:conclusions} concludes on the present work and its contributions.

\section{Related work}
\label{sec:related-work}

\noindent{\bf Temporal segmentation of videos and photostreams.\quad} Classical approaches for temporal segmentation were built for videos on hand-crafted features aiming at capturing visual image content \cite{koprinska2001temporal}. Current state of the art approaches \cite{xu2015discriminative,theodoridis2016multi} use as an intermediate representation semantic features that are more invariant with respect to abrupt visual changes in the field of view. When the videos are captured by a camera worn on the head that hence moves with the wearer, motion based features have proved to be specially useful \cite{poleg2014temporal}. 
However, in the domain of egocentric photostreams, motion information  is not available due to the low frame rate.
Therefore, Tavalera et al.~\cite{talavera2015r} focused on a new temporal segmentation algorithm based on graph-cuts and used global image features extracted through a pre-trained CNN for representing each frame. Later on, \cite{dimiccoli2017sr} improved this framework by adding a semantic level to the feature representations of egocentric photostreams. In particular, a semantic vocabulary of concepts was computed and used in addition to contextual features, where the concept scores are the confidence of the occurrence of the concepts in each frame. 
Paci et al.~\cite{paci2016context} proposed a similarity learning approach based on Siamese ConvNets that aims at learning a similarity function between low-resolution egocentric images.
Recently, Dias and Dimiccoli \cite{dias2019learning} have proposed to learn event representations in a fully unsupervised fashion by predicting the temporal context. Specifically, they proposed a neural network model and an LSTM model performing a self-supervised pretext task consisting in predicting the concept vectors of neighbor frames given  the  concept  vector  of  the current frame. This work has shown the importance of encoding the temporal context to improve event representations.  A similar approach to learn feature representations with LSTM networks is proposed in \cite{del2018predicting}. 
%
Yet, unlike \cite{dias2019learning}, in which the different models are learnt on-the-fly and unsupervised for single image sequences, \cite{del2018predicting} relies on a huge dataset for training the LSTM model in an unsupervised way.

\noindent{\bf Nonlocal self-similarity.\quad} 
However, in these works, the temporal context of a frame is encoded only \emph{locally} by considering its neighbors.
In this paper, we built on the concept of \emph{nonlocal self-similarity} at temporal level to improve event representations by encoding nonlocal temporal context. 
The concept was first used in \cite{efros1999texture} and has found its most prominent application in the nonlocal means algorithm for image denoising \cite{buades2005non}. 
The underlying key idea is that for every small patch in an image, it is possible to find many similar patches in the same image 
(possibly after affine transformations); 
these can be used for denoising.
For nonlocal means denoising, the concept was extended to 1D time series in \cite{tracey2012nonlocal}.
Along a different line, Dimiccoli and Salembier \cite{Dimiccoli09monocular,dimiccoli2009hierarchical} proposed to exploit spatial nonlocal self-similarity for improving segmentation boundaries in images in the context of a hierarchical segmentation algorithm. This is achieved by modeling each pixel by its probability distribution conditioned to those of neighbor pixels. Doing so, boundary pixels are typically put together before being grouped to the object they belong to, hence ensuring the boundary smoothness.

Here, we extend this idea to \emph{temporal segmentation}. To the best of our knowledge, nonlocal self-similarity has never been used to improve the segmentation of time series.

\section{Methodology}
\label{sec:methodology}

\subsection{Temporal nonlocal self-similarity}
\begin{figure}[t]
     \centering
     \includegraphics[width=0.48\textwidth]{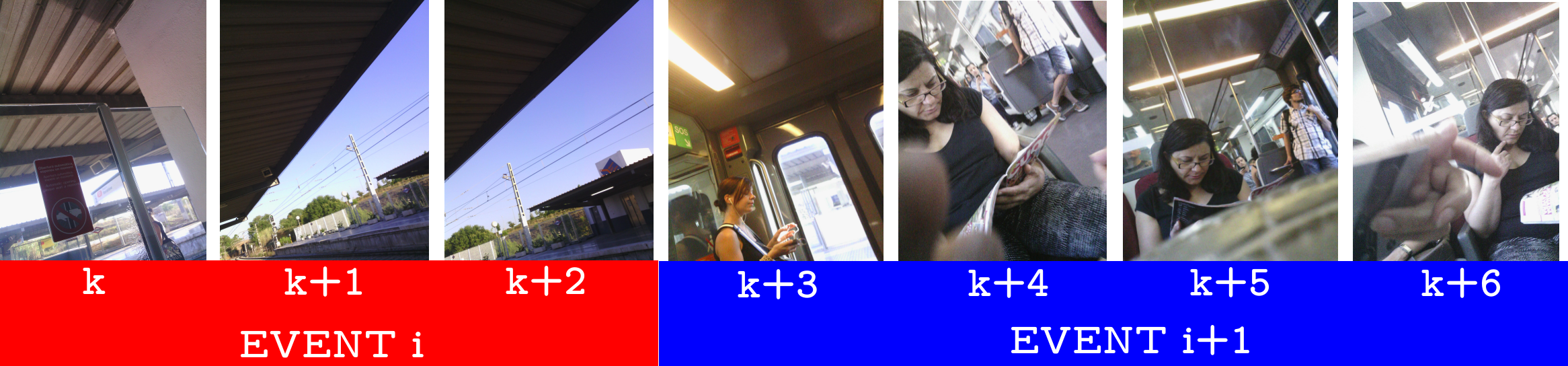}
     \caption{\label{fig:example}Examples of temporally adjacent events in an egocentric photostream, from the EDUB-Seg dataset \cite{dimiccoli2017sr}.}
\label{fig:more-egocentric-images}
\end{figure}

\noindent{\bf Model assumptions and intuitions.\quad} The key assumption used in our frame modeling is that an egocentric photostream can be considered as a fairly general stationary random process, meaning that, as the length of the photostream grows, for every small temporal segment in the sequence, it is possible to find many similar temporal segments in the same sequence. This is intuitively true when looking at the \emph{semantic representations} of small temporal segments, rather than the temporal segments themselves. For instance, all small temporal segments with people in a train or bus will have similar semantic features (such as appearance of a person, neon, etc. \cite{dimiccoli2017sr}), and the same is true for all small temporal segments captured while walking in the street, etc., while the images themselves can typically be very different, cf. Fig. \ref{fig:example} for an example.
\\%
\noindent{\bf Self-similarity function.\quad} 
The \emph{self-similarity function} is designed to quantify the similarity between frames  of a temporal segment centered at $k$ and a temporal segment centered at $j$, and is defined as follows.
Let $u(k)\in \mathbb{R}^P$ denote a vector of $P$ image features at time $k$, $k=1,\ldots,K$, where $K$ is the length of the sequence. Further, let $\mathcal{N}_k=\{k-M,\ldots,k-1,k+1,\ldots,k+M\}$ denote the indices of the $2M$ neighboring feature vectors of $u(k)$.
In analogy with 2D data (images) \cite{efros1999texture,buades2005non,dimiccoli2009hierarchical}, 
the self-similarity function of  $u(k)$ in a temporal sequence, conditioned to
its temporal neighborhood $\mathcal{N}_k$, is given by the quantity:
\begin{equation}
S^{NL}(k,j) = \frac{1}{\mathcal{Z}(k)} \exp \left(-\frac{ d(u(\mathcal{N}_k), u(\mathcal{N}_j))} {h}\right),
\end{equation}
where $d(u(\mathcal{N}_k),u(\mathcal{N}_j)) = \sum_{i=1}^{2M}   ||u(\mathcal{N}_k(i))-u(\mathcal{N}_j(i))||^2$ is the sum of the Euclidean distances of the vectors in the neighborhoods of $k$ and $j$,
$\mathcal{Z}(k)$ is a normalizing factor such that $\sum_jS^{NL}(k,j)=1$, ensuring that $S^{NL}(k,j)$ can be interpreted as a conditional probability of $u(j)$ given $u(\mathcal{N}_k)$, as detailed in \cite{buades2005non}, and $h$ is the parameter that tunes the decay of the exponential function. Below, $h$ is fixed such that the median of $\mathcal{Z}(k)\cdot S^{NL}(k,j)$ over all couples $(k,j)$ equals $\frac{1}{2}$.
%
%
\\
\noindent{\bf Nonlocal self-similarity features.\quad} 
The key idea in the proposed temporal segmentation approach is to use the (dis)similarity between a frame $k$ and other frames $j$ in the photostream, quantified by $S^{NL}(k,j)$, as a feature for temporal segmentation.
In other words, we model each frame $k$ by its associated self-similarity function $S^{NL}(k,j)$: we replace the set of \emph{local} features $u(k)$ with a new set of \emph{nonlocal} features 
\begin{equation}
u^{NL}(k) = \{S^{NL}(k,j)\}_{j=k\pm 1, 2, \ldots} \in \mathbb{R}^{N},
\end{equation}
where $N$ is the size of the temporal  interval where self-similarity is computed. In the experiments reported below, we use \emph{all} other frames of the sequence in a full nonlocal fashion, i.e., $N=K-1$. 
Under the model assumptions, the similarity of $u^{NL}(k)$ and $u^{NL}(k')$ will be large if $k$ and $k'$ belong to the same event, and it will be small if $k$ and $k'$ belong to two different neighboring events, and this property will be exploited for temporal segmentation.

\subsection{Temporal segmentation algorithm}
To compute the temporal segmentation, we employed the same algorithm used in \cite{dias2019learning}. It is based on building a structured representation of a set of hierarchical partitions in which the finest level of detail is given by the initial partition of all frames. The nodes of
the tree are associated to frames that represent the union of two children frames and the root node represents the entire image sequence. The tree is constructed in an ascending hierarchical fashion, in which the two temporally neighboring nodes with smallest distance between each other are united to form a new node. It is important to note that the algorithm is constrained to join only \emph{neighboring} nodes to form a new node, in contradistinction with classical hierarchical clustering \cite{dias2019learning}. The union of the frames of each node is modeled as the average over the frames associated with the node, and the Euclidean norm is used as a distance between two nodes. The algorithm is here applied to the main principal components of the frames $u$ or $u^{NL}$, after standardization over the entire sequence of frames.

\begin{table}[t]
\centering
\setlength{\tabcolsep}{2.2pt}
\noindent\begin{tabular}{|l||r|r|r|r|r|r|r|}
\hline 
 & base & NNF & NNFB & NNFB & NNFB & NNFB & LSTM\\
& line& $n=1$& $n=2$& $n=3$& $n=4$& $n=5$& $n=1$\\
\hline 
\hline 
L & $0.46$ & $0.50$ & $0.54$ & $0.51$ & \boldmath$0.56$ & $0.49$ & $0.53$\\
NL & \boldmath$0.58$ & \boldmath$0.52$ & \boldmath$0.59$ & \boldmath$0.54$ & $0.52$ & \boldmath$0.54$ & \boldmath$0.56$\\\hline
Diff. & $+0.12$ & $+0.03$ & $+0.05$ & $+0.04$ & $-0.05$ & $+0.04$ & $+0.03$\\
\hline 
\end{tabular}
\caption{\label{tab:avresults}{\bf Average temporal segmentation performance.} F-measure for temporal segmentation on 7 different sets of local features (L) and on their nonlocal self-similarity (NL), averaged over users (best results marked in bold).}
\end{table}

\begin{table}[t]
\centering
\setlength{\tabcolsep}{1.3pt}
\noindent\begin{tabular}{|l|l||r|r|r|r|r|r|r|}
\hline 
& & base & NNF & NNFB & NNFB & NNFB & NNFB & LSTM\\
&& line& $n=1$& $n=2$& $n=3$& $n=4$& $n=5$& $n=1$\\
\hline 
User&L & $0.31$ & $0.43$ & $0.51$ & $0.52$ & \boldmath$0.55$ & $0.43$ & $0.50$\\
1-1&NL & \boldmath\color{red}$0.65$ & \boldmath$0.61$ & \boldmath$0.57$ & \boldmath$0.56$ & $0.48$ & \boldmath$0.51$ & \boldmath$0.54$\\
\hline 
User&L & $0.36$ & \boldmath$0.38$ & \boldmath$0.55$ & $0.33$ & \boldmath$0.42$ & \boldmath$0.33$ & \boldmath$0.52$\\
1-2&NL & \boldmath\color{red}$0.56$ & $0.35$ & $0.42$ & \boldmath$0.35$ & $0.33$ & $0.32$ & $0.42$\\
\hline 
User&L & $0.46$ & \boldmath$0.63$ & $0.63$ & $0.62$ & $0.64$ & $0.57$ & $0.61$\\
1-3&NL & \boldmath\color{red}$0.87$ & $0.56$ & \boldmath$0.80$ & \boldmath$0.72$ & $0.64$ & \boldmath$0.69$ & \boldmath$0.69$\\
\hline 
User&L & $0.50$ & \boldmath$0.54$ & $0.54$ & $0.51$ & \boldmath$0.61$ & $0.56$ & $0.45$\\
2-1&NL &\boldmath $0.55$ & $0.50$ & \boldmath$0.57$ & \boldmath$0.59$ & $0.51$ & \boldmath\color{red}$0.61$ & \boldmath$0.56$\\
\hline 
User&L & $0.65$ & $0.67$ & $0.71$ & $0.64$ & $0.70$ & \boldmath$0.73$ & $0.67$\\
2-2&NL & \boldmath$0.70$ & \boldmath\color{red}$0.78$ & \boldmath$0.75$ & \boldmath$0.75$ & $0.70$ & $0.66$ & \boldmath$0.75$\\
\hline 
User&L & $0.68$ & \boldmath$0.78$ & \boldmath$0.79$ & $0.78$ & $0.78$ & $0.72$ & $0.79$\\
2-3&NL & \boldmath$0.78$ & $0.75$ & $0.71$ & $0.78$ & \boldmath\color{red}$0.85$ & \boldmath$0.80$ & \boldmath\color{red}$0.85$\\
\hline 
User&L & $0.43$ & \boldmath$0.45$ & \boldmath$0.40$ & $0.35$ & $0.40$ & $0.41$ & $0.40$\\
3-1&NL & \boldmath$0.45$ & $0.39$ & $0.35$ & \boldmath$0.46$ & \boldmath$0.43$ &\boldmath $0.43$ & \boldmath\color{red}$0.49$\\
\hline 
User&L & \boldmath$0.40$ & $0.31$ & $0.34$ & \boldmath$0.43$ & \boldmath$0.47$ & \boldmath$0.37$ & $0.37$\\
3-2&NL & $0.25$ & \boldmath$0.40$ & \boldmath\color{red}$0.63$ & $0.20$ & $0.33$ & $0.36$ & \boldmath$0.39$\\
\hline 
User&L & $0.41$ & $0.40$ & $0.40$ & $0.43$ & \boldmath\color{red}$0.59$ & $0.36$ & \boldmath$0.47$\\
4&NL & \boldmath$0.42$ & $0.40$ & \boldmath$0.50$ & \boldmath$0.47$ & $0.32$ & \boldmath$0.42$ & $0.46$\\
\hline 
User&L & $0.35$ & $0.39$ & $0.49$ & $0.44$ & $0.49$ & $0.44$ & \boldmath$0.48$\\
5&NL & \boldmath$0.56$ & \boldmath$0.49$ & \boldmath$0.55$ & \boldmath$0.53$ & \boldmath$0.58$ & \boldmath\color{red}$0.59$ & $0.44$\\
\hline 
\end{tabular}
\caption{\label{tab:results}{\bf Temporal segmentation performance per user.} 
F-measure for temporal segmentation on 7 different sets of local features (L) and on their nonlocal self-similarity (NL) for each users. Best results per user and feature are marked in bold, best results for each user in red.}
\end{table}

\begin{figure*}[t]
\begin{minipage}[t]{0.33\linewidth}
\centering
\includegraphics[width=\linewidth]{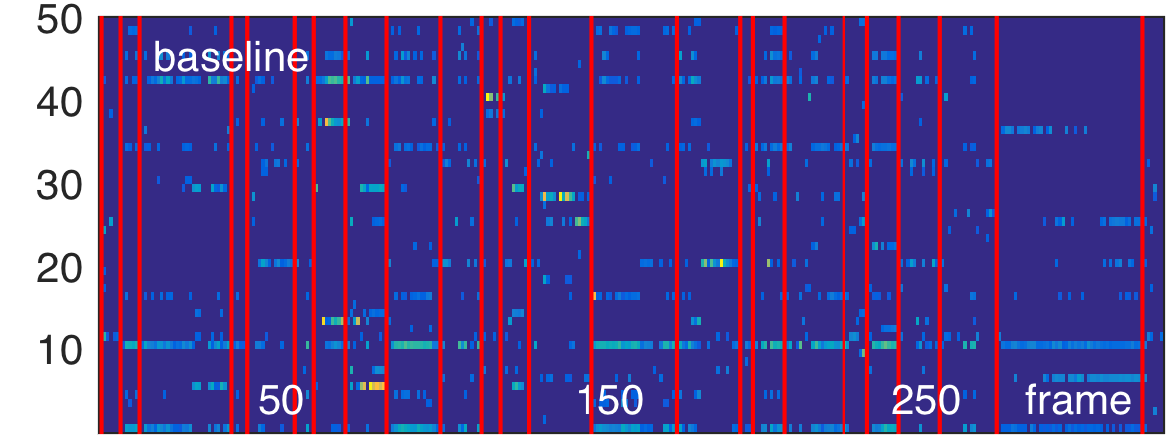}\\
\includegraphics[width=\linewidth]{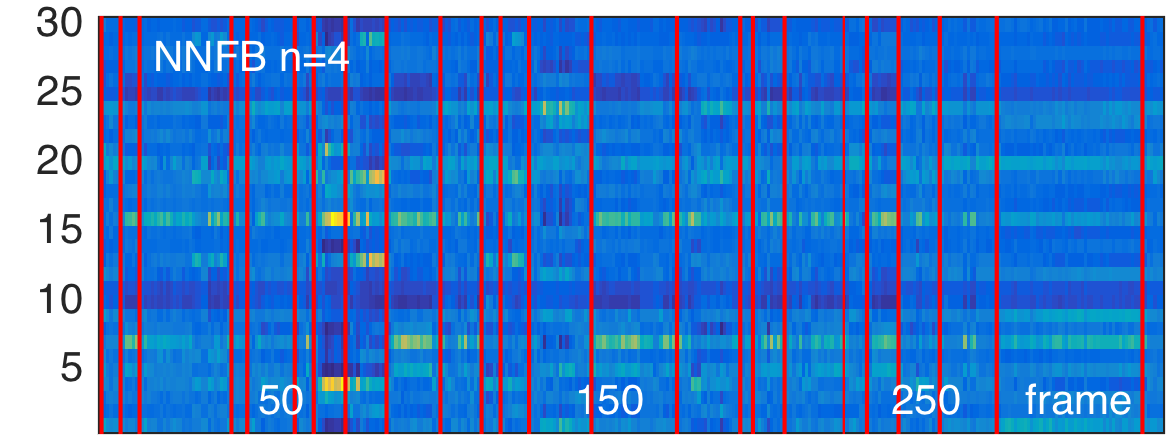}\\
\includegraphics[width=\linewidth]{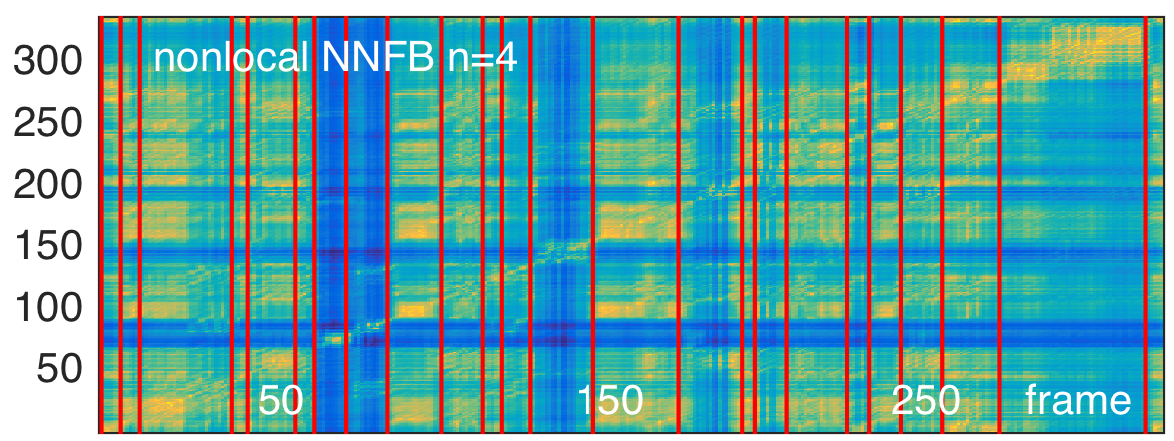}\vskip-1.8mm
\quad (a)
\end{minipage}%
\begin{minipage}[t]{0.33\linewidth}
\centering
\includegraphics[width=\linewidth]{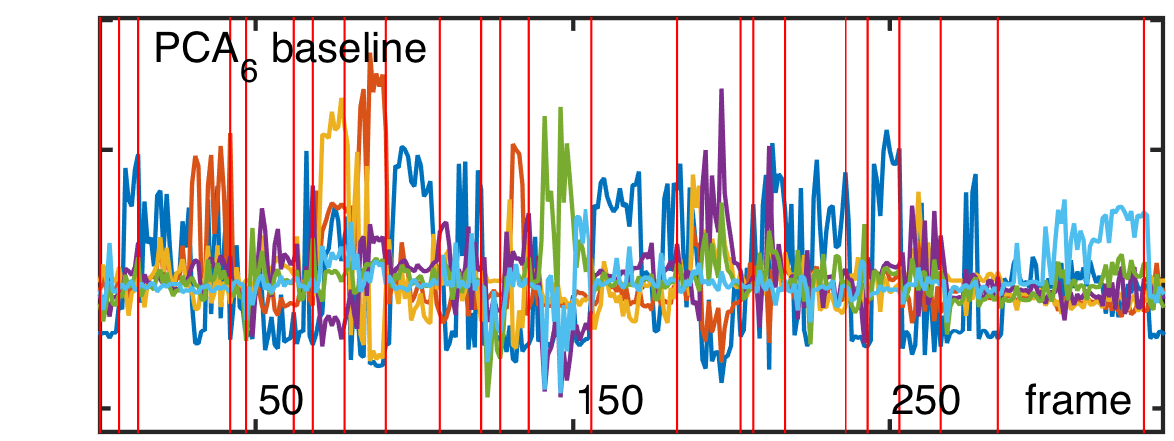}\\
\includegraphics[width=\linewidth]{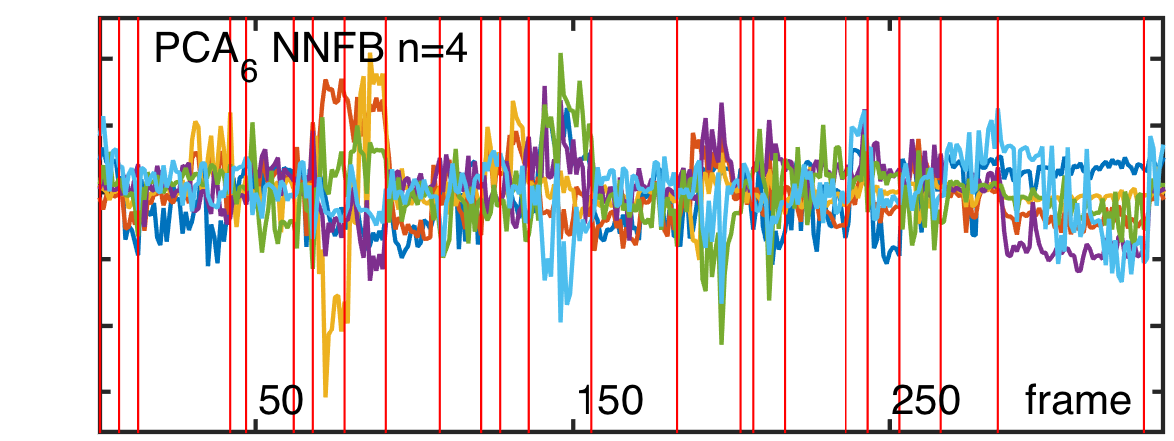}\\
\includegraphics[width=\linewidth]{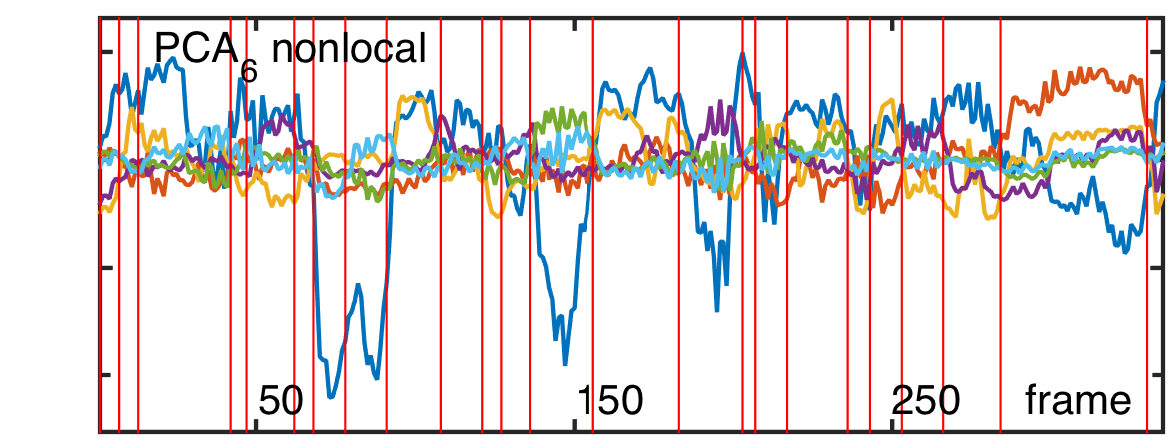}\vskip-1.8mm
\quad (b)
\end{minipage}%
\begin{minipage}[t]{0.33\linewidth}
\centering
\includegraphics[width=\linewidth]{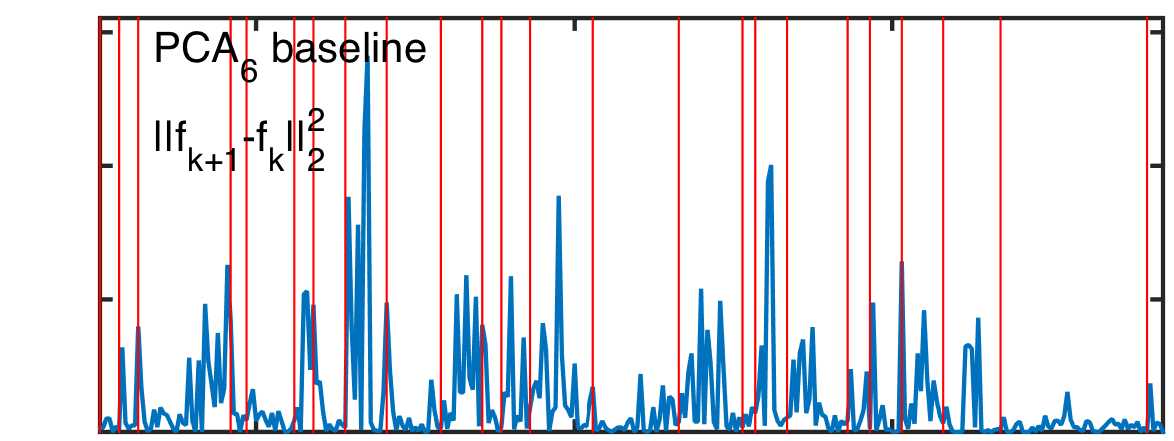}\\
\includegraphics[width=\linewidth]{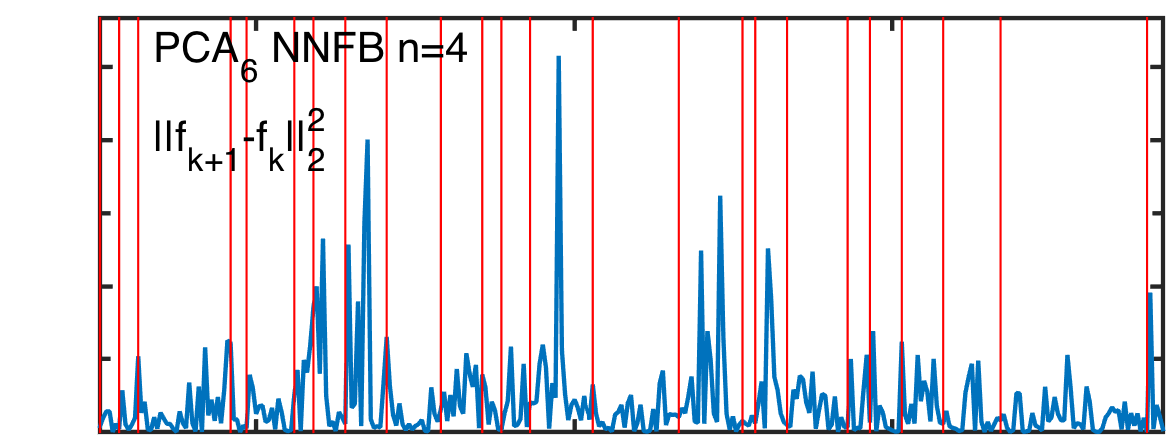}\\
\includegraphics[width=\linewidth]{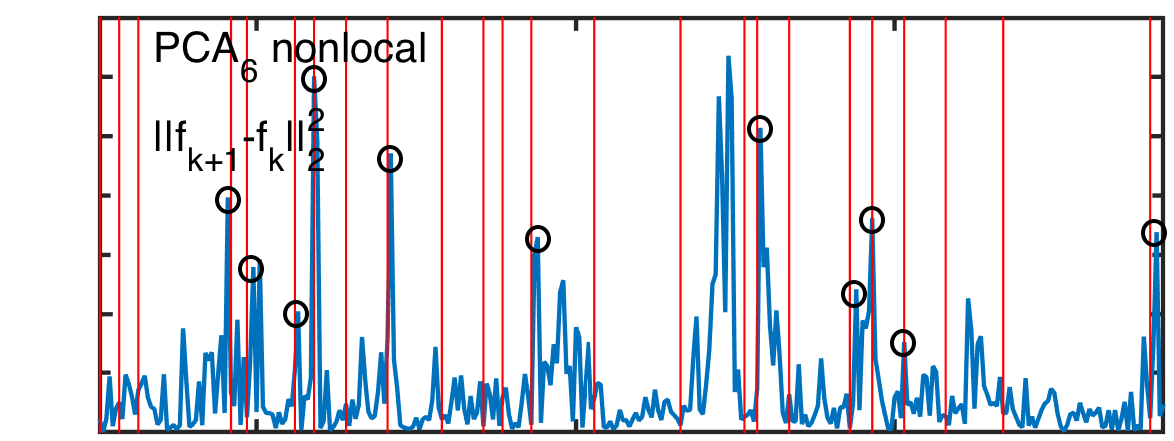}\vskip-1.8mm
\quad (c)
\end{minipage}%
\caption{\label{fig:feature}{\bf Event segmentation for User2-3.}
(a) Baseline features (top panel), learnt NNFB $n=4$ features (center panel) and corresponding nonlocal self-similarity (bottom panel)  for User2-3; (b) corresponding main principal components; (c) L2 difference between neighboring frames. Vertical red bars indicate ground truth events.}
\end{figure*}

\section{Temporal segmentation results}
\label{sec:results}

\subsection{Dataset and features}
\label{sec:data}
\noindent{\bf Dataset and performance evaluation.\quad}
We used a subset of the EDUB-Seg dataset  as in \cite{dimiccoli2017sr,dias2019learning} consisting of ten
image  sequences for  five  different  users,  captured  by  a  wearable
photo-camera  that  takes two pictures per minute,  with an average of 662 images per sequence. This subset comes together with  the  ground  truth  event  segmentation  and  concept  vectors  describing  the
probability of each concept in the image, cf. \cite{dias2019learning} for details. 
The event segmentation performance for the EDUB-Seg dataset is quantified using the F-measure, calculated for a tolerance of $\pm5$ frames as in \cite{talavera2015r,dimiccoli2017sr,paci2016context,dias2019learning,del2018predicting}.
The number of temporal segments
is set for each sequence separately such as to yield maximal F-measure for the sequence.

\noindent{\bf Local features and nonlocal self-similarity.\quad}
Here we used different features extracted from the images in \cite{dias2019learning}: CNN based features consisting of indicator vectors for concepts detected in the images \cite{dimiccoli2017sr}, denoted baseline, and six sets of features embedding the local temporal context obtained in \cite{dias2019learning} by, respectively, a simple feed-forward NN (NNF), forward-backward auto-encoding NNs for different temporal depths $n=1,2,3,4$, and an LSTM auto-encoder, cf.~\cite{dias2019learning} for details. For each of these sets of features $u(k)$, we compute the 6 main principal components and use them as local features for event segmentation (\cite{dias2019learning} did not use PCA).
Further, we compute the nonlocal self-similarity features $u^{NL}(k)$, as defined in Sec.~\ref{sec:methodology}, for these local features, extract the 6 main principal components and use them as nonlocal features for temporal segmentation.
The temporal patch size for computing $u^{NL}(k)$ was set to $\pm 2$ frames (i.e., $M=2$).

\subsection{Temporal segmentation performance}
\label{sec:performance}
\noindent{\bf Illustration for a single user.\quad}
Fig.~\ref{fig:feature} (a) plots the baseline features (top panel), and NNFB $n=4$ (center panel) and corresponding nonlocal features $u^{NL}$ (bottom panel) for User2-3. Subplot (b) reports the temporal evolution of the corresponding 6 main principal components of these features, and in subplot (c) the Euclidean distance between neighboring frames is quantified. 
 Visual comparison of the panels in Fig.~\ref{fig:feature} (a) indicates that baseline features suffer from quite large and abrupt sporadic changes in feature values within segments, making robust temporal segmentation difficult. While the features with local temporal context (NNF, NNFB, LSTM) improve upon this situation and display less within-event variability, their nonlocal self-similarity provides a more coherent picture of temporal evolution and clearly yields a more solid basis for temporal segmentation. Inspection of the main principal components in Fig.~\ref{fig:feature} (b) confirms this observation: baseline features display strong erratic fluctuations; features with local temporal context are still somewhat noisy; nonlocal features yield a cleaner temporal evolution. Finally, using the Euclidean norm of the difference between frames as an indicator for event boundaries, Fig.~\ref{fig:feature} (c) indicates that the improved robustness of the nonlocal features pays off in terms of temporal segmentation accuracy: indeed, a larger number of dissimilarity peaks get lined up with true event boundaries (as indicated by the black circles).

\noindent{\bf Average temporal segmentation performance.\quad}
The average F-measure of the temporal segmentations for all users are reported in Tab.~\ref{tab:avresults}  for the different local features and the corresponding nonlocal self-similarity features.
The results unambiguously demonstrate that the proposed use of nonlocal self-similarity is beneficial and clearly improves the temporal segmentation performance. The nonlocal features yield F-measures larger than $0.5$. The use of nonlocal features is particularly beneficial for the baseline features that are unaware of temporal context (the F-measure is increased by $0.12$). Yet, also for the features encoding local temporal context (NNF, NNFB, LSTM), nonlocal self-similarity leads to significant (though smaller) performance improvements.

\noindent{\bf Temporal segmentation performance per user.\quad}
Tab.~\ref{tab:results} provides a detailed view and reports the F-measures obtained for each individual user.
%
It has already been observed in \cite{dimiccoli2017sr} that there is a relatively large variability in event segmentation quality for the different users of the EDUB-Seg dataset. Nevertheless, if we do not consider User1-2 and User3-2 for which temporal segmentation performance is poor overall, with F-measure values $\ll0.5$, the results indicate that nonlocal features lead to better temporal segmentations also for each of the user individually. Moreover, the best F-measure value obtained for the different features for each individual user (in red in Tab.~\ref{tab:results}) is consistently obtained by nonlocal features.
%

To conclude, these results clearly indicate that the use of nonlocal self-similarity benefits the temporal segmentation of image sequences and leads to improved performance.

\section{Conclusions}
\label{sec:conclusions}
This paper contributed to the problem of temporal segmentation, which is recognized to be crucial for several computer vision tasks.
Temporal segmentation performance are tightly coupled with the underlying feature representation, and previous work showed the importance of encoding local temporal context.
Here, we focused on enhancing the discriminative power of feature representations using temporal context \emph{nonlocally}. This is achieved in an original way by building on the concept of nonlocal self-similarity.
We validated our approach on the popular EDUB-Seg dataset, showing that 
the proposed method
leads to a consistent improvement with respect to state of the art feature representations, be they aware or not of the local temporal context.
In future work, we will explore how to learn event representations by leveraging the nonlocal self-similarity principle within a deep learning framework. 

\clearpage

\bibliographystyle{IEEEbib}
\bibliography{mibib}

\begin{thebibliography}{10}

\bibitem{zelnik2001event}
Lihi Zelnik-Manor and Michal Irani,
\newblock ``Event-based analysis of video,''
\newblock in {\em Proc. IEEE CVPR}, 2001, vol.~2, pp. II--II.

\bibitem{liwicki2015online}
Stephan Liwicki, Stefanos~P Zafeiriou, and Maja Pantic,
\newblock ``Online kernel slow feature analysis for temporal video segmentation
  and tracking,''
\newblock {\em IEEE Transactions on Image Processing}, vol. 24, no. 10, pp.
  2955--2970, 2015.

\bibitem{poleg2014temporal}
Yair Poleg, Chetan Arora, and Shmuel Peleg,
\newblock ``Temporal segmentation of egocentric videos,''
\newblock in {\em Proc. IEEE CVPR}, 2014, pp. 2537--2544.

\bibitem{dimiccoli2017sr}
Mariella Dimiccoli, Marc Bola{\~n}os, Estefania Talavera, Maedeh Aghaei,
  Stavri~G Nikolov, and Petia Radeva,
\newblock ``Sr-clustering: Semantic regularized clustering for egocentric photo
  streams segmentation,''
\newblock {\em Computer Vision and Image Understanding}, vol. 155, pp. 55--69,
  2017.

\bibitem{del2018predicting}
Ana~Garcia del Molino, Joo-Hwee Lim, and Ah-Hwee Tan,
\newblock ``Predicting visual context for unsupervised event segmentation in
  continuous photo-streams,''
\newblock {\em arXiv preprint arXiv:1808.02289}, 2018.

\bibitem{spriggs2009temporal}
Ekaterina~H Spriggs, Fernando De~La~Torre, and Martial Hebert,
\newblock ``Temporal segmentation and activity classification from first-person
  sensing,''
\newblock in {\em Proc. IEEE CVPRW}, 2009, pp. 17--24.

\bibitem{kruger2017efficient}
Bj{\"o}rn Kr{\"u}ger, Anna V{\"o}gele, Tobias Willig, Angela Yao, Reinhard
  Klein, and Andreas Weber,
\newblock ``Efficient unsupervised temporal segmentation of motion data,''
\newblock {\em IEEE Transactions on Multimedia}, vol. 19, no. 4, pp. 797--812,
  2017.

\bibitem{kennedy2010detecting}
Robert~E Kennedy, Zhiqiang Yang, and Warren~B Cohen,
\newblock ``Detecting trends in forest disturbance and recovery using yearly
  landsat time series: 1. landtrendr—temporal segmentation algorithms,''
\newblock {\em Remote Sensing of Environment}, vol. 114, no. 12, pp.
  2897--2910, 2010.

\bibitem{jamali2015detecting}
Sadegh Jamali, Per J{\"o}nsson, Lars Eklundh, Jonas Ard{\"o}, and Jonathan
  Seaquist,
\newblock ``Detecting changes in vegetation trends using time series
  segmentation,''
\newblock {\em Remote Sensing of Environment}, vol. 156, pp. 182--195, 2015.

\bibitem{adams2016hierarchical}
Roy Adams, Nazir Saleheen, Edison Thomaz, Abhinav Parate, Santosh Kumar, and
  Benjamin Marlin,
\newblock ``Hierarchical span-based conditional random fields for labeling and
  segmenting events in wearable sensor data streams,''
\newblock in {\em ICML}, 2016, pp. 334--343.

\bibitem{bari2018rconverse}
Rummana Bari, Roy~J Adams, Md~Mahbubur Rahman, Megan~Battles Parsons, Eugene~H
  Buder, and Santosh Kumar,
\newblock ``rconverse: Moment by moment conversation detection using a mobile
  respiration sensor,''
\newblock {\em Proceedings of the ACM on Interactive, Mobile, Wearable and
  Ubiquitous Technologies}, vol. 2, no. 1, pp. 2, 2018.

\bibitem{koprinska2001temporal}
Irena Koprinska and Sergio Carrato,
\newblock ``Temporal video segmentation: A survey,''
\newblock {\em Signal processing: Image communication}, vol. 16, no. 5, pp.
  477--500, 2001.

\bibitem{abdulhussain2018methods}
Sadiq Abdulhussain, Abd Ramli, M~Saripan, Basheera Mahmmod, Syed Al-Haddad, and
  Wissam Jassim,
\newblock ``Methods and challenges in shot boundary detection: a review,''
\newblock {\em Entropy}, vol. 20, no. 4, pp. 214, 2018.

\bibitem{paci2016context}
Francesco Paci, Lorenzo Baraldi, Giuseppe Serra, Rita Cucchiara, and Luca
  Benini,
\newblock ``Context change detection for an ultra-low power low-resolution
  ego-vision imager,''
\newblock in {\em Proc. ECCV Workshops}, 2016, pp. 589--602.

\bibitem{dias2019learning}
Catarina Dias and Mariella Dimiccoli,
\newblock ``Learning event representations by encoding the temporal context,''
\newblock in {\em Proc. ECCV Workshops}, Cham, 2019, pp. 587--596, Springer
  International Publishing.

\bibitem{xu2015discriminative}
Zhongwen Xu, Yi~Yang, and Alex~G Hauptmann,
\newblock ``A discriminative cnn video representation for event detection,''
\newblock in {\em Proc. IEEE CVPR}, 2015, pp. 1798--1807.

\bibitem{theodoridis2016multi}
Thomas Theodoridis, Anastasios Tefas, and Ioannis Pitas,
\newblock ``Multi-view semantic temporal video segmentation,''
\newblock in {\em Proc. IEEE ICIP}, 2016, pp. 3947--3951.

\bibitem{talavera2015r}
Estefania Talavera, Mariella Dimiccoli, Marc Bolanos, Maedeh Aghaei, and Petia
  Radeva,
\newblock ``R-clustering for egocentric video segmentation,''
\newblock in {\em Iberian Conference on Pattern Recognition and Image
  Analysis}. Springer, 2015, pp. 327--336.

\bibitem{efros1999texture}
Alexei~A Efros and Thomas~K Leung,
\newblock ``Texture synthesis by non-parametric sampling,''
\newblock in {\em Proc. IEEE ICCV}, 1999, p. 1033.

\bibitem{buades2005non}
Antoni Buades, Bartomeu Coll, and J-M Morel,
\newblock ``A non-local algorithm for image denoising,''
\newblock in {\em Proc. IEEE CVPR}, 2005, vol.~2, pp. 60--65.

\bibitem{tracey2012nonlocal}
Brian~H Tracey and Eric~L Miller,
\newblock ``Nonlocal means denoising of ecg signals,''
\newblock {\em IEEE Transactions on Biomedical Engineering}, vol. 59, no. 9,
  pp. 2383--2386, 2012.

\bibitem{Dimiccoli09monocular}
Mariella Dimiccoli,
\newblock {\em Monocular depth estimation for image segmentation and
  filtering},
\newblock Phd thesis, Technical University of Catalonia (UPC), 2009.

\bibitem{dimiccoli2009hierarchical}
Mariella Dimiccoli and Philippe Salembier,
\newblock ``Hierarchical region-based representation for segmentation and
  filtering with depth in single images.,''
\newblock in {\em Proc. IEEE ICIP}, 2009, pp. 3533--3536.

\end{thebibliography}

\end{document}